%% file: Paperdp.tex
\documentclass[a4paper,art14]{article}
\usepackage{epsfig}
\usepackage{graphics}
\begin{document}

\title{
\textbf{The trilinear neutral Higgs self-couplings  \\ in  the MSSM. \\
Complete one-loop analysis}}
\author{ M. V. Dolgopolov $^{a}$ \footnote{E-mail: dolg@ssu.samara.ru} , Yu. P.
Philippov $^{a}$ \footnote{E-mail: yufilberg@pochta.ru}  \\
[0.5cm]
\small $^{a}$ Department of general and theoretical physics, \\
\small Samara State University, Samara, Russia}
\date{}
\maketitle

\vspace{-1 cm}
\begin{abstract}
The Higgs boson effective self-couplings $\lambda_{hhh}$ and
$\lambda_{HHH}$ are calculated in
the framework of the Minimal Supersymmetric Standard Model (MSSM)
for the complete set of one-loop diagrams with one of the Higgs bosons
off-shell. The comparison with previous results, where only
the leading correction terms in the limiting case of large masses
of virtual particles were calculated, is carried out. We analyse
the dependence of the self-couplings on the energy and ${\tt tan} \beta$;
it is demonstrated that the tree-level self-couplings could acquire
substantial one-loop corrections, which could be
phenomenologically important.
\end{abstract}

\input Intro1
\input Section11
\input Section21
\input Section31
\input Section41
\input Conclusion
\input Acknowledge

\input Literat
\end{document}

%% file: Intro1.tex
\section*{Introduction}

\hspace{5mm} Two basic elements of the gauge boson and fermion
mass generation in the Standard Model (SM) and its minimal
supersymmetric extension (MSSM) are (1) the spontaneous symmetry
breaking (SSB) \cite{SBS} and (2) the Higgs mechanism
\cite{Higgsmeh}. Gauge bosons and fermions gain masses via
gauge-invariant interaction with scalar fields,  which have
non-zero vacuum expectation values. The SSB and Higgs mechanism
respect gauge invariance and renormalizability of the model. In
order to confirm experimentally the mechanism of mass generation,
it is needed (1) to detect the Higgs bosons and measure their
masses (2) to measure the Higgs boson couplings with gauge bosons
and fermions (3) to determine the Higgs boson self-couplings. The
last step is important especially for the case of MSSM, where the
self-couplings are determined by the soft supersymmetry breaking
principle. Tree-level analysis shows that some neutral Higgs boson
self-couplings can be measured \cite{Zer00, D-K-M-Z, Daw99} with
relatively high accuracy on the future high-luminosity colliders.
The leading one-loop corrections to the neutral Higgs
self-couplings may be expected to be sizable in some regions of
MSSM parameter space, with the potential to change substantially
the tree-level results.

We calculate two self-couplings of the neutral MSSM Higgs bosons
$h^0$, $H^0$ taking into account the complete set of one-loop
diagrams. Our main interest is the analysis of modifications in
the Higgs potential by the corrections coming from the scalar and
sparticle sectors of the model.

In section 1 the Higgs boson interaction Lagrangian and the
self-couplings are presented. In sections 2 and 3 we discuss
technical details of the approach. Some numerical results and
their comparative analysis are contained in section 4.

%% file: Section11.tex
\section{Higgs boson interaction Lagrangian}

\hspace{5mm} The Higgs sector of MSSM includes five physical
fields:

$\diamond$ two neutral $CP-$ even Higgs fields $\{h^0,H^0\}$;

$\diamond$ neutral  $CP-$ odd Higgs field $A^0$;

$\diamond$ charged fields $ H^\pm$.

The Higgs boson interaction Lagrangian has the form

\begin{equation}
 \mathcal{L}_{Int}^{Higgs}=
 \mathcal{L}_{Int}^{(3)}+\mathcal{L}_{Int}^{(4)}
\end{equation}

$\mathcal{L}_{Int}^{(3)}-$  Lagrangian of the triple Higgs boson
interactions, \newline $\mathcal{L}_{Int}^{(4)}-$ Lagrangian of
the quartic Higgs  boson interactions.

\begin{eqnarray}
\mathcal{L}_{Int}^{(3)}=
\frac{\lambda_{hhh}}{3!}hhh+\frac{\lambda_{hhH}}{2!}hhH+
\frac{\lambda_{hHH}}{2!}hHH+ \nonumber \\
\frac{\lambda_{HHH}}{3!}HHH+ \frac{\lambda_{hAA}}{2!}hAA+
\frac{\lambda_{HAA}}{2!}HAA+ \nonumber \\
\lambda_{hH^+H^-}hH^+H^- +\lambda_{HH^+H^-}HH^+H^-
\label{eq:TripleInt}
\end{eqnarray}
The tree level Higgs boson self-couplings are represented in
(\ref{eq:TripleInt}) as functions of the two free parameters: mass
of
 Higgs field $A^0$, $M_A$ and mixing angle $\beta$.
 \begin{eqnarray}
\lambda_{hhh}&=&-3ia_0\cos(2\alpha)\sin(\alpha+\beta)
\label{eq:lhhh}\\
\lambda_{hhH}&=&-ia_0 [2\sin(2\alpha)\sin(\alpha+\beta) -
\cos(2\alpha)\cos(\alpha+\beta)]
\label{eq:lhhH}\\
\lambda_{hHH}&=&ia_0 [2\sin(2\alpha)\cos(\alpha+\beta)+
\cos(2\alpha) \sin(\alpha+\beta)]
\label{eq:lhHH}\\
\lambda_{HHH}&=&-3ia_0\cos(2\alpha)\cos(\alpha+\beta)
\label{eq:lHHH}\\
\lambda_{hAA}&=&-ia_0\cos(2\beta)\sin(\alpha+\beta)
\label{eq:lhAA}\\
\lambda_{HAA}&=&ia_0\cos(2\beta)\cos(\alpha+\beta)
\label{eq:lhAA}\\
\lambda_{hH^+H^-}&=&-igM_W
\sin(\beta-\alpha)-ia_0\cos(2\beta)\sin(\alpha+\beta)
\label{eq:lhHpHp}\\
\lambda_{HH^+H^-}&=&-igM_W \cos(\beta-\alpha)
-ia_0\cos(2\beta)\cos(\alpha+\beta) \label{eq:lhHpHp}
\end{eqnarray}
where $M_W$, $M_Z$ - gauge boson masses, $\theta$ - Weinberg
angle, $g$ - $SU(2)_L$ gauge constant. Connections between the
mixing angles $\alpha$ and $\beta$ and also the parameter $a_0$
are determined by the conditions
\begin{eqnarray}
\tan 2\alpha= \tan 2\beta \frac{M_A^2+M_Z^2}{M_A^2-M_Z^2} ,
\hspace{4mm}
 a_0=\frac{igM_z}{2\cos(\theta_W)}
\end{eqnarray}

%% file: Section21.tex
\section{Perturbation theory. Vertex function}

\hspace{5mm} Main object of our study is the vertex function (VF)
of the triple Higgs interaction.

\begin{figure}[t!]
\begin{center}
\includegraphics{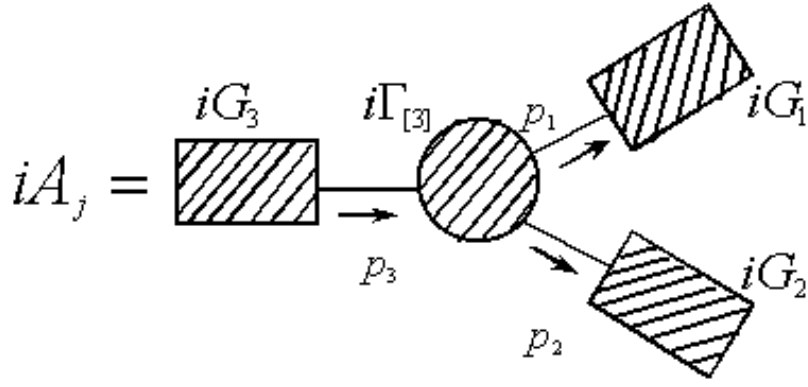}
\end{center}
\vspace{-3mm}
 \caption{{Feynman diagram for amplitude $A_j$.}}
\label{fig:VerFun}
\end{figure}

Following the standard normalization conventions \cite{Bogol-Shir,
Lund-Liv}, the amplitude of the given diagram can be represented
as follows (see Fig. \ref{fig:VerFun}):

\begin{equation}
-iA_j=-iG_1(p_1,\{a_1\})(i\Gamma_{[3]}(p_1,p_2,p_3,\{c\}))
G_2(p_2,\{a_2\}) G_3(p_3,\{a_3\}) \label{eq:AmpFey}
\end{equation}
where $G_i(p_i,\{a_i\})$ -- the Green's functions beyond the given
diagram, $p_i$\,--\, the four-vector of $i-th$ Higgs boson.
$\{a\},\{c\}$\,--\,sets of model parameters, characterizing given
diagram. $\Gamma_{[3]}(p_1,p_2,p_3,\{c\})$ --\,{\it vertex
function of the triple Higgs interaction}. It can be represented
as
\begin{equation}
\Gamma_{[3]}(p_1,p_2,p_3,\{c\})=\sum_{l=0}^{\infty}
\Gamma_{[3]}^{(l)}(p_1,p_2,p_3,\{c\}) \label{eq:SumGam}
\end{equation}

In the last expression the decomposition in $\alpha_e$ is
performed. Each term coincides with the sum of one-particle
irreducible diagrams of the order of $\alpha_e$.  Our analysis is
carried out for the decomposition
\begin{equation}
\Gamma_{[3]}(p_1,p_2,p_3,\{c\})=\Gamma_{[3]}^{(0)}(p_1,p_2,p_3,\{c\})+
\Gamma_{[3]}^{(1)}(p_1,p_2,p_3,\{c\}) \label{eq:1loopGam}
\end{equation}

\begin{figure}[t!]
\begin{center}
\includegraphics{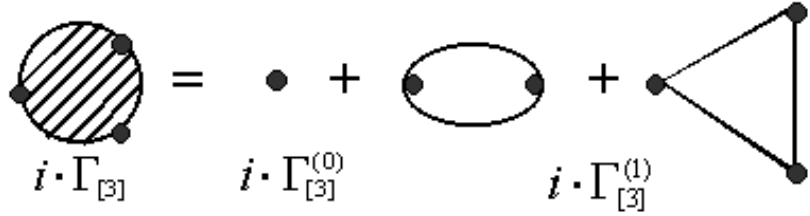}
\end{center}
\vspace{-3mm}
 \caption{Vertex function decomposition: the one-loop
approximation} \label{fig:razlogenie1}
\end{figure}

Graphical interpretation of the last formula is shown in Fig.
\ref{fig:razlogenie1}.

The condition (\ref{eq:SumGam}) in terms of self-couplings has the
form
\begin{equation}
\lambda_{Hg_1Hg_2Hg_3}=\lambda_{Hg_1Hg_2Hg_3}^{(0)}+
\Delta\lambda_{Hg_1Hg_2Hg_3}^{(1)} =\lambda_{Hg_1Hg_2Hg_3}^{(0)}
(1+\alpha_e \Delta\tilde{\lambda}_{Hg_1Hg_2Hg_3}^{(1)})
\label{eq:1looplam}
\end{equation}

The complete one-loop contribution (Feynman gauge) is determined
by contributions of physical fields (standard fermions, sfermions,
gauge bosons, chargino, neutralino), as well as Goldstone modes
and gost fields.
\begin{eqnarray}
\Delta\lambda_{Hg_1\,Hg_2\,Hg_3}^{(1)}=\Delta\lambda^{(1)}(f)+
\Delta\lambda^{(1)}(\tilde{f})+\Delta\lambda^{(1)}(Gb)+
\Delta\lambda^{(1)}(Ch)+\nonumber \\
\Delta\lambda^{(1)}(Neu)+\Delta\lambda^{(1)}(Hg)+
\Delta\lambda^{(1)}(Gs)+\Delta\lambda^{(1)}(Gh)
\label{eq:1looplam}
\end{eqnarray}

The following analysis is based on the general formulas for the
one-loop contributions to two- and three-point vertex functions.
Our algorithms are implemented in the computer algebra programs
that allow to operate with bulky intermediate expansions.

%% file: Section31.tex
\section{Scheme of calculations}

\hspace{5mm} Main features of our calculation are

1. The center-of-mass frame is used for explicit phase space
formulas,
   when one Higgs boson is a virtual particle.

2. Feynman gauge was used.

3. The solutions of renormalization group equations for gauge
   constants and third
   generation quarks masses  were
   employed to take into account the energy dependence
   more precisely \cite{H-H-H}.

4. Tensor reduction for the scalar one-loop integrals was
   used \cite{TernsorRed}.

5. On-shell renormalization procedure was employed
\cite{Ch-P-R-D}.

The experimental data were taken from \cite{physparam}.

%% file: Section41.tex
\section{Results and their analysis}

\hspace{5mm} Two types of dependences are represented in this
paper: Higgs self-couplings dependences on $\tan \beta$,
$\sqrt{s}$.
\begin{figure}[t!]
\hspace{-7mm}
\includegraphics{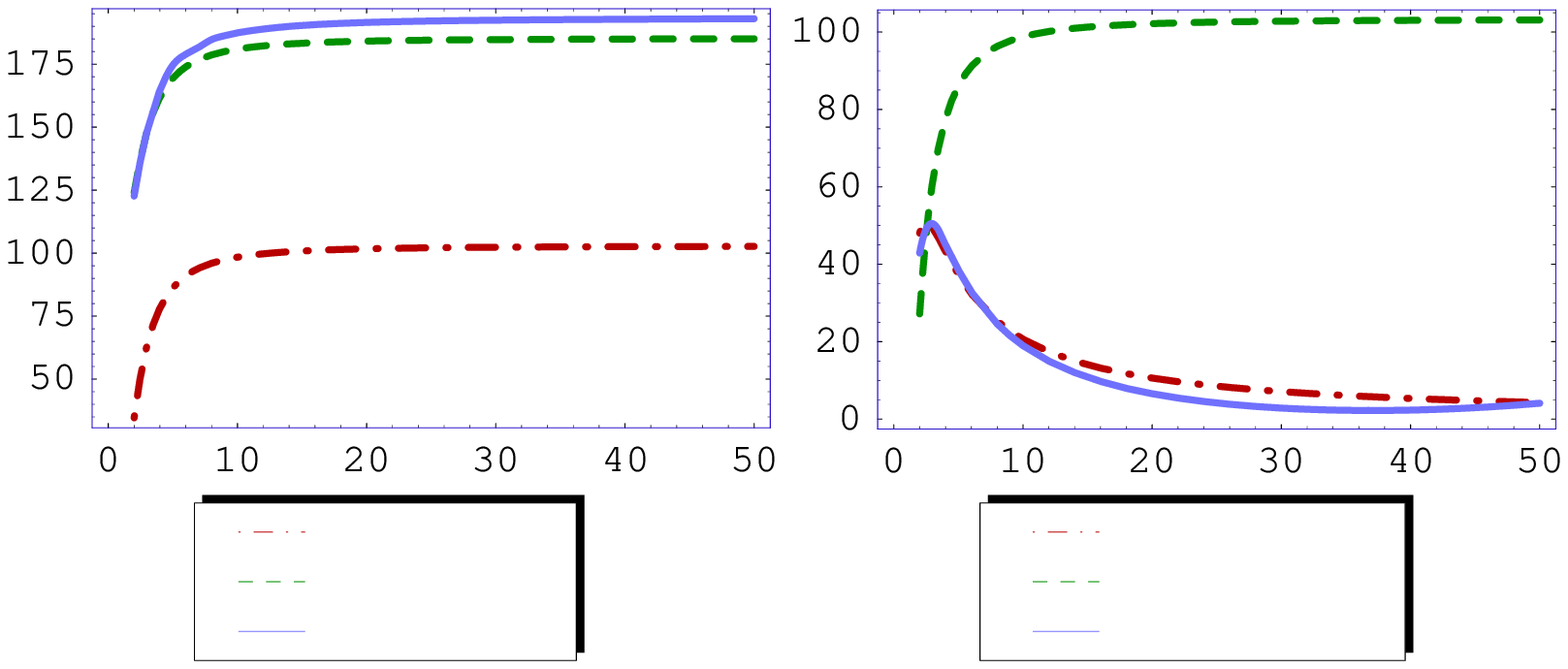}
\put(-450,30){\mbox{\bf a)}} \put(-215,30){\mbox{\bf b)}}
\put(-290,50){\mbox{\bf \large $\tan \beta$}}
\put(-50,50){\mbox{\bf  \large $\tan \beta$}}
\put(-490,210){\mbox{\bf \large $\lambda_{hhh},[GeV] $ }}
\put(-260,210){\mbox{\bf  \large $\lambda_{HHH},[GeV] $}}
\put(-390,47){\mbox{\bf \normalsize $\lambda_{hhh}^0$}}
\put(-390,32){\mbox{\bf \normalsize $\lambda_{hhh}^0+\Delta
\lambda_{hhh}^{\scriptsize \textrm{Lead}}$}}
\put(-390,17){\mbox{\bf \normalsize
 $\lambda_{hhh}^0+\Delta \lambda_{hhh}^{\scriptsize
\textrm{Comp}}$}} \put(-155,47){\mbox{\bf \normalsize
$\lambda_{HHH}^0$}} \put(-155,32){\mbox{\bf \normalsize
$\lambda_{HHH}^0+\Delta \lambda_{HHH}^{\scriptsize
\textrm{Lead}}$}} \put(-155,17){\mbox{\bf \normalsize
 $\lambda_{HHH}^0+\Delta \lambda_{HHH}^{\scriptsize
\textrm{Comp}}$}} \put(-260,210){\mbox{\bf \large
$\lambda_{HHH},[GeV] $}} \vspace{-4mm} \caption{ Dependence of the
trilinear Higgs self-couplings
    (on the  tree and one-loop levels) on  $\tan \beta$, for
    $\sqrt{s}=m_{Hg_1}+m_{Hg_2}+10$ GeV, $M_A=500$ GeV, $M_3=\mu=300$ GeV,
$M_{\tilde{Q}}=M_{\tilde{U}}=M_{\tilde{D}}=
M_{\tilde{R}}=M_{\tilde{L}}=1$ TeV, $A_f=200$ GeV.}
\label{fig:GraphArray1}
\end{figure}
\begin{figure}[t!]
\hspace{-4mm}
\includegraphics{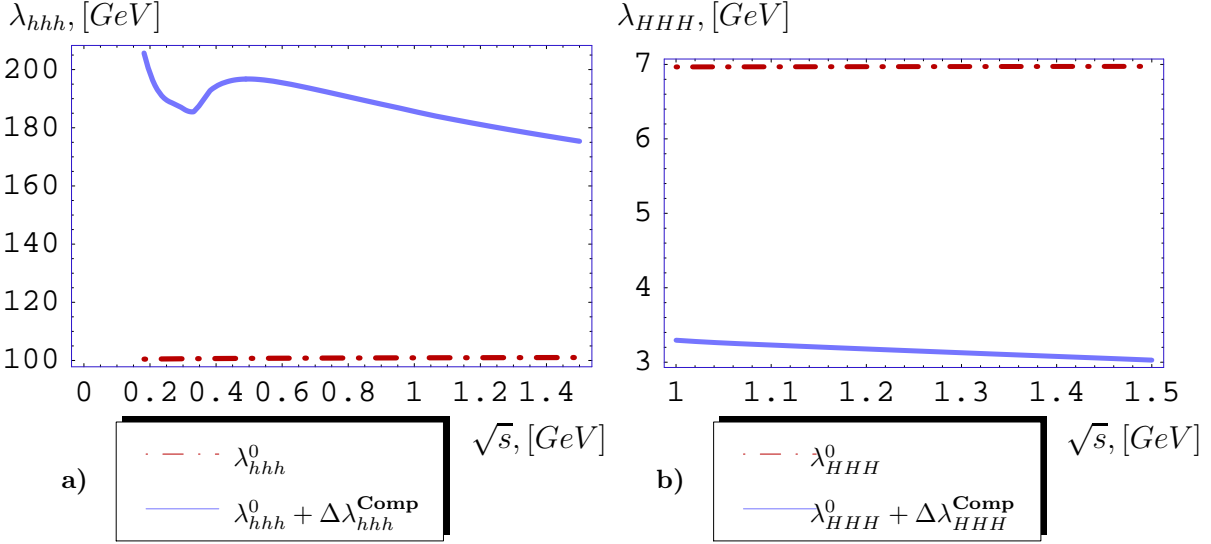}
\put(-450,30){\mbox{\bf a)}} \put(-225,30){\mbox{\bf b)}}
\put(-295,45){\mbox{\bf \large $\sqrt{s}, [GeV]$}}
\put(-70,45){\mbox{\bf \large $\sqrt{s}, [GeV]$}}
\put(-470,205){\mbox{\bf \large $\lambda_{hhh},[GeV] $ }}
\put(-240,205){\mbox{\bf  \large $\lambda_{HHH},[GeV] $}}
\put(-385,38){\mbox{\bf  \normalsize $\lambda_{hhh}^0$}}
\put(-385,17){\mbox{\bf \normalsize
 $\lambda_{hhh}^0+\Delta \lambda_{hhh}^{\scriptsize
\textrm{Comp}}$}} \put(-167,38){\mbox{\bf \normalsize
$\lambda_{HHH}^0$}} \put(-167,17){\mbox{\bf \normalsize
$\lambda_{HHH}^0+\Delta \lambda_{HHH}^{\scriptsize
\textrm{Comp}}$}} \vspace{-4mm} \caption{Dependence of the
trilinear Higgs self-couplings
    (on the tree and one-loop levels) on  $\sqrt{s}$, for $\tan \beta=30$,
      $M_A=500$ GeV, $M_3=\mu=300$ GeV, $M_{\tilde{Q}}=M_{\tilde{U}}=M_{\tilde{D}}=
      M_{\tilde{R}}=M_{\tilde{L}}=1$ TeV, $A_f=200$ GeV.}
\label{fig:GraphArray2}
\end{figure}
In Fig. \ref{fig:GraphArray1} three kinds of dependences on $\tan
\beta$ are represented: tree-level self-couplings ($\lambda^{0}$),
self-couplings  with leading $t-\tilde{t}$ one-loop correction
($\lambda^{0}+\Delta \lambda^{Lead}$) and self-couplings with
complete one-loop contribution ($\lambda^{0}+\Delta
\lambda^{Comp}$). Analytic expressions for ($\Delta
\lambda^{Lead}$) can be found in \cite{oneloopcor} in the case of
large mass approximation for virtual particles.

Obviously, differences between ($\lambda_{hhh}^{0}+\Delta
\lambda_{hhh}^{Lead}$) and ($\lambda_{hhh}^{0}+\Delta
\lambda_{hhh}^{Comp}$) are insignificant (9.5 \% from curve
($\lambda_{hhh}^{0}+\Delta \lambda_{hhh}^{Lead}$) for $\tan \beta
=45$). It is important to note that the sizable value of
correction does not contradict the perturbation theory
applicability.

In \cite{Hollik} it was shown that the self-coupling
($\lambda_{hhh}^{0}+\Delta \lambda_{hhh}^{Comp}$) in the
decoupling limit (our situation for $M_A=500$ GeV satisfy this
limit) can be represented in terms of Higgs boson $h^0$ mass. By
redefinition of ($\lambda_{hhh}^{0}+\Delta \lambda_{hhh}^{Lead}$)
the correction value is being insignificant and do not exceed
$6\%$. The analogous conditions for ($\lambda_{hhh}^{0}+\Delta
\lambda_{hhh}^{Comp}$) give the correction of $11\%$. Another
situation is observed in $\lambda_{HHH}-$ case. The curve
coinciding to ($\lambda_{HHH}^{0}+\Delta \lambda_{HHH}^{Lead}$)
demonstrates huge one-loop leading contribution in
$\lambda_{HHH}$, however the complete  one-loop contribution is
very small. The reason of this discrepancy is in the usage of the
large masse approximation for virtual particles. We have not used
any approximations for one-loop scalar integrals. The
applicability of the large masse approximation is from our point
of view questionable in the situation under consideration (maybe
except the $hhh$ - case) because of too large masses of virtual
Higgs bosons.

In Fig. \ref{fig:GraphArray2} the self-couplings dependences on
the process energy passing by virtual Higgs boson are represented.
One can observe that $\lambda_{HHH}$ is very weekly dependent on
$\sqrt{s}$. For a given situation a future high-luminosity
colliders
 can not detect these dependences because we can certainly
 suppose that $\lambda_{HHH}$ is not a running parameter. Another
 situation is observed in $\lambda_{hhh}-$ case. For instance for
 $\sqrt{s}=186$ GeV the self-coupling $\lambda_{hhh}=205$ GeV and for
 $\sqrt{s}=1500$ GeV the self-coupling $\lambda_{hhh}=175$ GeV. If
the accuracy of the self-coupling measurement on the future
colliders will be better than 30 GeV, that given dependence could
be detected.

The summing error of results is determined by determination error
of $t-$ quark mass ($\delta m = 5.1 $ GeV) and by generality of
soft supersymmetry breaking parameters values. Errors for
$\lambda_{hhh}-$ case are equal 5 GeV and 8 GeV, and for
$\lambda_{HHH}-$ case 1 GeV and 2 GeV accordingly (for $M_A=500$
GeV).

%% file: Conclusion.tex
\section*{Conclusion}

\hspace{5mm} In this work  the neutral Higgs boson self-couplings
with complete set of the one-loop diagrams are represented and
analyzed in the framework of the MSSM. The effective dependences
of the Higgs boson self-couplings $\lambda_{hhh}$, $\lambda_{HHH}$
on $\tan \beta$ and energy $\sqrt{s}$ are evaluated  with
 one of the Higgs bosons is off-shell. The authors have
avoided any approximations in one-loop scalar integrals
calculations. It has been shown, that the tree-level
self-couplings can acquire sizable one-loop corrections. These
 must be taken into account for detailed comparative analysis of theory
and experimental data, which could be phenomenologically
important.

%% file: Acknowledge.tex
\section*{Acknowledgements}

\hspace{5mm} The authors thank E.E. Boos, M.N. Dubinin and  A.A.
Biryukov for the valuable discussions and useful critical remarks.

%% file: Paperdp.bbl
\begin{thebibliography}{1}
\addcontentsline{toc}{section}{}
\bibitem{SBS} J. Goldstone - Nuovo Cimento, \textbf{v19} 154, (1961);
              Y. Nambu, G. Jona-Lasinio - Phys. Rev., \textbf{v122} 345,(1961);
              J. Goldstone, A. Salam, S. Weinberg - Phys. Rev., \textbf{v127}
              965,(1962);
\bibitem{Higgsmeh} P.W. Higgs Phys. Lett., \textbf{12} 132 (1964);
                   E. Englert, R. Brout Phys. Rev. Lett., \textbf{13} 321 (1964);
                   P.W. Higgs Phys. Rev. Lett., \textbf{13} 508 (1964);
                   G.S. Guralnik, C.R. Hagen, T.W.B. Kibble Phys. Rev. Lett.,
                   \textbf{13} 585 (1964);
                   P.W. Higgs Phys. Rev., \textbf{v145} 1156 (1966);
                   T.W.B. Kibble Phys. Rev., \textbf{v155} 1554 (1967);
\bibitem{Zer00} R.M. Zerwas, hep-ph/0003221.
\bibitem{D-K-M-Z} A. ~Djouadi, W. Kilian, M. Muhlleitner, P.M. Zerwas,
hep-ph/0001169, DESY 99/171.
\bibitem{Daw99} S.Dawson, hep-ph/9912433.
\bibitem{oneloopcor}  H.E. Haber, and R. Hempfling, Phys. Rev. Lett. \textbf{66}
 1815, (1991); Y. Okada, M. Yamaguchi, T. Yanagida, Prog. Theor. Phys. \textbf{85}
 1, (1991); J. Ellis, G. Ridolfi and F. Zwirner, Phys. Lett. \textbf{B257} 83, (1991).

V.Barger, M.S. Berger, A.L. Stange and R.J.N. Phillips, Phys. Rev.
\textbf{D45} 4128, (1992); Z.Kunszt and F. Zwirner, Nucl. Phys.
\textbf{B385} 3, (1992) .

A. Djouadi, H.E. Haber, P.M. Zerwas, Phys. Lett. \textbf{B375}
203, (1996).
\bibitem{Bogol-Shir}
By N.N. Bogolyubov, D.V. Shirkov, Introduction to the theory of
quantized fields. (Steklov Math. Inst., Moscow $\&$ Dubna, JINR).
Published in Intersci.Monogr.Phys.Astron. \textbf{3} 1,(1959).
\bibitem{Lund-Liv}
V. B. Berestetskii, L.D. Landau, E.M. Lifshitz, L. P. Pitaevskii,
Quantum Electrodynamics, 2nd ed. Oxford, England: Pergamon Press,
(1982).
\bibitem{H-H-H} H.E. Haber, R. Hempfling, A.H. Hoang,
                hep-ph/9609331.
\bibitem{TernsorRed}  L.M. Brown, R.P. Feynman, Phys.Rev., \textbf{85} 231,
                      (1952);
                      G. Passarino, M.J.G. Veltman, Nucl.Phys., \textbf{B160} 151,
                      (1979).
\bibitem{Ch-P-R-D}
P.H. Chankovski, S. Pokorski, J. Rosiek, Nucl. Phys.,
\textbf{B423} 437, (1994); A. Dabelstein, Z. Phys., \textbf{C67}
495, (1995), Nucl. Phys., \textbf{B456} 25, (1995).

\bibitem{physparam} LEP EWWG, http://lepewwg.web.cern.ch/LEPEWWG/plots
/summer2000/;\\
D.E.Groom et. al, Eur. Phys. J. \textbf{c15} 1, (2000);\\
S. Bethke, J. Phys. \textbf{G26 R27}, (2000);\\
A.D. Martin, J.Outhwaite, M.C. Ryskin, Phys. Lett. \textbf{B492}
69, (2000).
\bibitem{Hollik} W.Hollik, S. Penaranda, Eur.Phys.J.\textbf{C23} 163, (2002).
\end{thebibliography}
